\documentclass[conference]{IEEEtran}
\usepackage{tabularx}
\usepackage{cite}
\usepackage{amsmath,amssymb,amsfonts}
\usepackage{algorithmic}
\usepackage{graphicx}
\usepackage{textcomp}
\usepackage[dvipsnames]{xcolor}
\usepackage{booktabs}
\usepackage{adjustbox}
\usepackage{caption}
\usepackage{float}
\usepackage{multirow}
\usepackage{siunitx}
\usepackage{xspace}
\usepackage{hyperref} 
\newcommand{\manualdataset}{{\textsc{JavaVFC}}\xspace}
\newcommand{\autodataset}{{\textsc{JavaVFC-extended}}\xspace}

\title{\manualdataset: Java Vulnerability Fixing Commits from Open-source Software}

\usepackage{fancyhdr}
\begin{document}

\setcounter{page}{1}

\author{
    \IEEEauthorblockN{Tan Bui\IEEEauthorrefmark{1},
        Yan Naing Tun\IEEEauthorrefmark{1},
        Yiran Cheng\IEEEauthorrefmark{2},
        Ivana Clairine Irsan\IEEEauthorrefmark{1},
        Ting Zhang\IEEEauthorrefmark{1},
        Hong Jin Kang\IEEEauthorrefmark{1}
        }
        
    \IEEEauthorblockA{\IEEEauthorrefmark{1}School of Computing and Information Systems, Singapore Management University, Singapore}
    
    \IEEEauthorblockA{\IEEEauthorrefmark{2}Institute of Information Engineering, Chinese Academy of Sciences, Beijing, China}
    
    \IEEEauthorblockA{\{ngoctanbui, yannaingtun, ivanairsan, tingzhang.2019, hjkang.2018\}@smu.edu.sg}

    \IEEEauthorblockA{chengyiran@iie.ac.cn}
}

\maketitle

\begin{abstract}
We present a comprehensive dataset of Java vulnerability-fixing commits (VFCs) to advance research in Java vulnerability analysis. Our dataset, derived from thousands of open-source Java projects on GitHub, comprises two variants: \manualdataset and \autodataset. The dataset was constructed through a rigorous process involving heuristic rules and multiple rounds of manual labeling. We initially used keywords to filter candidate VFCs based on commit messages, then refined this keyword set through iterative manual labeling. The final labeling round achieved a precision score of 0.7 among three annotators. We applied the refined keyword set to 34,321 open-source Java repositories with over 50 GitHub stars, resulting in \manualdataset with 784 manually verified VFCs and \autodataset with 16,837 automatically identified VFCs. Both variants are presented in a standardized JSONL format for easy access and analysis. This dataset supports various research endeavors, including VFC identification, fine-grained vulnerability detection, and automated vulnerability repair. The \manualdataset and \autodataset are publicly available at \href{https://zenodo.org/records/13731781}{https://zenodo.org/records/13731781}.

\end{abstract}

\section{Introduction}
In recent years, software vulnerabilities have garnered increasing attention within the software engineering community. Numerous studies have focused on understanding and automatically detecting these vulnerabilities~\cite{li2018vuldeepecker,chakraborty2021deep,li2021sysevr}. 
\textit{Data-driven} approaches to vulnerability detection and repair heavily rely on real-life vulnerability data for training and evaluation. However, a key challenge faced by these tools is the limited availability of high-quality datasets.

\vspace{8px}
\noindent\textbf{Motivation.} While various datasets have been created for vulnerability-related research, many of them suffer from significant limitations:
\begin{enumerate}
    \item Many existing datasets draw from reputable sources of vulnerability fixing commits (VFCs), such as the National Vulnerability Database\footnote{\href{https://nvd.nist.gov/}{https://nvd.nist.gov/}} (NVD) or CVEDetails\footnote{\href{https://www.cvedetails.com/}{https://www.cvedetails.com/}}~\cite{chen2023diversevul, fan2020ac, wang2023deepvd}. Despite their credibility, datasets based on CVE (Common Vulnerabilities and Exposures) data often lack in the number and diversity of VFCs they contain.
    \item The existing vulnerability datasets usually focus on C/C++~\cite{zheng2021d2a,fan2020ac}, leaving a gap when it comes to Java, one of the most widely used programming languages, especially in large-scale enterprise applications.
    \item Recent efforts to curate datasets including Java VFCs, such as the manually curated dataset from the \textit{SAP KB} project\footnote{\href{https://sap.github.io/project-kb/}{https://sap.github.io/project-kb/}}~\cite{ponta2019manually} and its extended dataset containing VFCs recorded in Mitre CVE database~\cite{zhou2021finding}, are limited in scope. They contain vulnerabilities from 205 and 310 open-source Java projects, respectively, with only 1,282 and 1,436 VFCs included.
\end{enumerate}

As a result, these datasets may not adequately represent the full spectrum of real-world scenarios or provide the comprehensive coverage necessary to train robust models. This underscores the pressing need for a new, high-quality VFC dataset that encompasses a larger and more diverse set of VFCs, particularly for Java projects.
To address these limitations, we provide a new dataset that specifically targets VFCs in Java projects on GitHub. By focusing on Java repositories, we aim to build a dataset that is both representative and relevant to the practical, real-world challenges faced by developers. Such a dataset would significantly enhance the development and evaluation of automated vulnerability detection and repair tools.

\vspace{8px}
\noindent\textbf{Contribution.} The contributions of this paper are two-fold:

\textbf{A. [Dataset:]} We present a comprehensive Java VFC dataset collected from a wide range of open-source Java projects. This dataset comes in two variants:
a) \manualdataset: A high-precision dataset comprising 784 VFCs, each meticulously verified by at least 2 out of 3 manual annotators.
b) \autodataset: A larger-scale dataset containing 16,837 VFCs, filtered using heuristics from 34,321 open-source Java projects.
Our dataset not only addresses a significant gap in the field but also lays the groundwork for future work in developing predictive models and tools to help developers identify and prioritize critical security fixes.

\textbf{B. [Keywords:]} We introduce a curated set of keywords designed to efficiently filter VFCs based on commit messages. 
This set of keywords facilitates future research in this area and can be easily extended.

These two contributions aim to enhance the resources available for vulnerability detection and repair in Java projects while also providing a framework for expanding efforts similar to those of other programming languages.

\section{Dataset Construction}

\begin{figure*}[tp]
    \centering
    \includegraphics[width=0.9\textwidth]{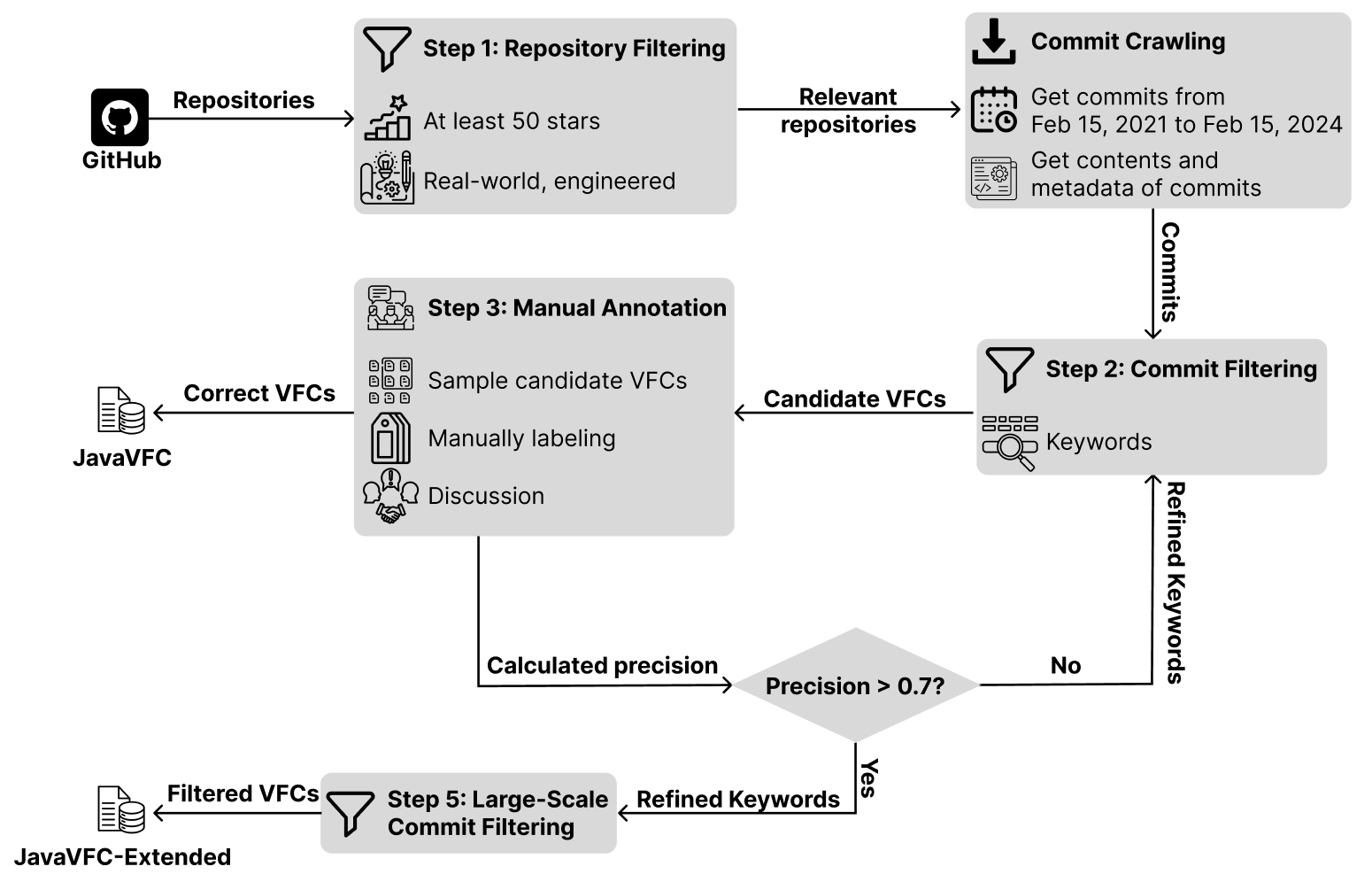}
    \caption{Data Collection Pipeline}
    \label{fig:workflow}
\end{figure*}

The data collection pipeline for \manualdataset is illustrated in Figure \ref{fig:workflow}. The process consists of five main steps:
\begin{enumerate}
    \item Step 1: Filtering GitHub repositories and downloading commits from February 15, 2021, to February 15, 2024.
    \item Step 2: Filtering commits to identify candidate VFCs.
    \item Step 3: Sampling candidate VFCs for manual annotation.
    \item Step 4: Iterating steps 2 and 3 to refine keywords and verify VFCs manually until achieving a precision of over 0.7.
    \item Step 5: Applying the final set of keywords to generate a larger pool of VFCs.
\end{enumerate}

\textbf{Evaluation Metrics.} We designate all commits filtered by keywords as ``candidate VFCs". 
From this set, we randomly sampled 1,900 candidate VFCs for manual annotation. 
During the annotation process, we evaluate two key metrics:
\begin{itemize}
    \item Precision: This metric assesses the proportion of candidate VFCs that are true VFCs. We calculate precision using Formula \ref{formula-p}, where TP (True Positives) represents candidate VFCs confirmed by human annotators as actual VFCs, and FP (False Positives) denotes those determined to be non-VFCs. This metric gauges the effectiveness of our keyword filtering in identifying true VFCs.
    \item Inter-annotator Agreement: We employ Fleiss' Kappa score to measure the consistency among multiple annotators~\cite{fleiss1981measurement}. This metric is computed using Formula \ref{formula-kappa}.
\end{itemize}

These metrics provide insights into both the accuracy of our keyword-based filtering and the reliability of our manual annotation process.

\begin{align}
\label{formula-p}
    \text{Precision} = \frac{\text{TP}}{\text{TP} + \text{FP}}
\end{align}

\begin{align}
\label{formula-kappa}
    \kappa &= \frac{\overline{P} - \overline{P_e}}{1 - \overline{P_e}} \\
    \text{where:} \nonumber \\
    \overline{P} &= \frac{1}{N} \sum_{i=1}^{N} P_i = \frac{1}{Nn(n-1)} \sum_{i=1}^{N} \left[ \sum_{j=1}^{k} n_{ij}(n_{ij} - 1) \right] \\
    \overline{P_e} &= \sum_{j=1}^{k} p_j^2 = \sum_{j=1}^{k} \left( \frac{1}{Nn} \sum_{i=1}^{N} n_{ij} \right)^2
\end{align}

\text{and:}

\( N \) is the number of items,
\( n \) is the number of raters per item,
\( k \) is the number of categories,
\( n_{ij} \) is the number of raters who assigned item \( i \) to category \( j \),
\( P_i \) is the agreement proportion for item \( i \), and
\( p_j \) is the proportion of all assignments to category \( j \).

\textbf{Step 1: Repository Filtering.} We utilized GitHub, the largest platform for open-source projects, to create a high-quality dataset for identifying VFCs. To ensure project quality, we first filtered for Java repositories with more than 50 stars. Next, to focus on engineered projects, we excluded repositories that were collections, tutorials, or not representative of real-world software development by examining their descriptions and removing those containing keywords like ``tutorial", ``interview", or ``course". After applying both filters, 34,321 repositories met these criteria. Once the repositories were cloned, we extracted commits from February 15, 2021, to February 15, 2024, saving metadata including commit hash, author, date, commit message, and detailed information about file and code changes.

\textbf{Step 2: Commit Filtering.} 
We initiated the commit filtering process using security-related keywords proposed by Zhou et al.~\cite{inproceedings}. Regular expressions were employed to search commit messages and filter out security-unrelated commits. The keyword set was refined iteratively over three rounds of manual annotation.

\begin{table}[t]
\centering
\begin{tabular}{llll}
\toprule
\textbf{Batch} & \textbf{\# Commits Evaluated} & \textbf{Kappa's Score} & \textbf{Precision} \\ 
% \midrule
\midrule
Round 1 & 400 & 0.37 & 41.00\% \\
\midrule
Round 2 & 1,100 & 0.47 & 39.36\% \\
\midrule
Round 3 & 400 & 0.74 & 71.50\% \\
\bottomrule
\end{tabular}
\caption{Kappa's Score and Precision across different rounds of commits}
\label{table:kappa_precision}
\end{table}

\textbf{Step 3: Manual Annotation.} For each sampled candidate VFC, three annotators independently labeled whether it was a true VFC. 
All annotators hold a minimum of a Bachelor's degree in Computer Science and possess at least five years of programming experience.
After computing the inter-rater agreement score, they discussed how to resolve disagreements and adjust keywords. Human-verified VFCs include those unanimously agreed upon during independent annotation and those classified as VFCs during discussion.
We conducted three rounds of labeling in total. Table~\ref{table:kappa_precision} presents Fleiss' Kappa score and precision for each round of manual checking.

\textbf{Step 5:} Finally, we applied the refined set of keywords to 34,321 open-source Java repositories to develop the \autodataset dataset. The updated keywords include additional patterns for identifying vulnerabilities.
Table~\ref{tab:regex_patterns} shows the different categories, and we present 5 keywords from each category as an example.
Additionally, non-relevant matches, such as MS-DOS or fixes related to test cases, were excluded using the patterns listed under the ``Exclusions" category in the same table.

\begin{table*}[t]
\centering
\small
\begin{adjustbox}{max width=\textwidth}
\begin{tabular}{lp{0.85\textwidth}}
\toprule
\textbf{Category} & \textbf{Example Patterns} \\ \midrule
\textbf{Denial of Service (DoS)} & 
\texttt{denial.of.service, ReDoS, dos, DOS.attack(s), infinite.loop} \\ \midrule
\textbf{Code Execution} & 
\texttt{remote.code.execution, RCE, buffer.overflow, use-after-free, directory.traversal} \\ \midrule
\textbf{Injection Attacks} & 
\texttt{XXE, XSS, HTML.injection, malicious.app.can.modify, SQL.injection} \\ \midrule
\textbf{Security Bypass} & 
\texttt{bypass.keyguard, unauthori[z|s]ed, fix.bypass, cross-origin, open.redirect} \\ \midrule
\textbf{Certificate Issues} & 
\texttt{lacking(?: proper)? certificate(s), certificate.revalidation, verify.holder, add.security.check, check.the.redirect.URI} \\ \midrule
\textbf{Other Security Vulnerabilities} & 
\texttt{vulnerability(?!.xml), session.fixation, clickjack, high.priority.security.fix, malicious} \\ \midrule
\textbf{Exclusions} & 
\texttt{redos, MS-DOS, ://stackoverflow.com, fix\textbackslash s+\textbackslash *?\textbackslash s*tests, log.\textbackslash w+Enabled} \\ \bottomrule
\end{tabular}
\end{adjustbox}
\caption{Categorized regex patterns with examples for vulnerability detection and exclusions}
\label{tab:regex_patterns}
\end{table*}

All data is stored in JSONL (JSON Lines) format, providing a convenient and efficient structure for large datasets. 
This format facilitates easy parsing and analysis across various programming environments.
The complete dataset is publicly available on \href{https://zenodo.org/records/13731781}{https://zenodo.org/records/13731781}.

\section{Dataset Description}
Our dataset provides comprehensive information about VFCs in Java open-source software (OSS) projects hosted on GitHub, covering the period from 15 February 2021 to 15 February 2024. 
% These datasets are meticulously curated to support research in security and vulnerability analysis, offering a rich set of attributes capturing various aspects of each commit. 
The VFCs are organized into JSONL files.
%\textcolor{red}{what is each file? Does each file correspond to one repository?}
The \texttt{javavfc.jsonl} file contains data from the \manualdataset dataset, while \texttt{javavfc\_extended.jsonl} includes data from the \autodataset dataset.
Additionally, we publish the regular expressions used for commit filtering in the \texttt{regex.txt} file, offering transparency into our filtering methodology.

Table \ref{tab:vfc_features} provides an overview of the features included in our datasets, along with their corresponding column names in the JSONL files and descriptions. These features encompass essential details such as the commit link, commit message, author name, commit date, files changed, and the diff of the commit. The commit link allows for easy access to the specific commit in the GitHub repository, enabling verification and further exploration. The commit message offers insights into the intent behind the commit, often reflecting the reasoning or context provided by the author. The author name identifies the individual who made the commit, which can be valuable for understanding the contribution patterns within a project. 
The commit date is presented in Unix Time Stamp format. This facilitates temporal analysis of the vulnerability fixes. 
The files changed and commit diff features show the specifics of the modifications made. 

\begin{table*}[t]
    \centering
    % \captionsetup{skip=10pt} % Adjust the gap size here
    % \begin{adjustbox}{max width=\textwidth}
    \begin{tabular}{llp{8cm}}
        \toprule
        \textbf{Features} & \textbf{Column Name in the JSONL} & \textbf{Description} \\ \midrule
        Commit Link & commit\_link & URL link to the specific commit in the GitHub repository \\ \midrule
        Commit Message & message & The message provided by the author to describe the purpose of the commit \\ \midrule
        Author Name & author & The name of the author who made the commit \\ \midrule
        Commit Date & date & The time in Unix Time Stamp format when the commit was made  \\ \midrule
        Files Changed & files & All the changed files \\ \midrule
        Commit Diff & diff\_raw & The diff of the commit showing the exact changes made to the code \\ 
        \bottomrule
    \end{tabular}
    % \end{adjustbox}
    \caption{Features, Column Names in the CSV, and Descriptions}
    \label{tab:vfc_features}
\end{table*}

\begin{figure*}[t]
    \centering
    \includegraphics[width=\textwidth]{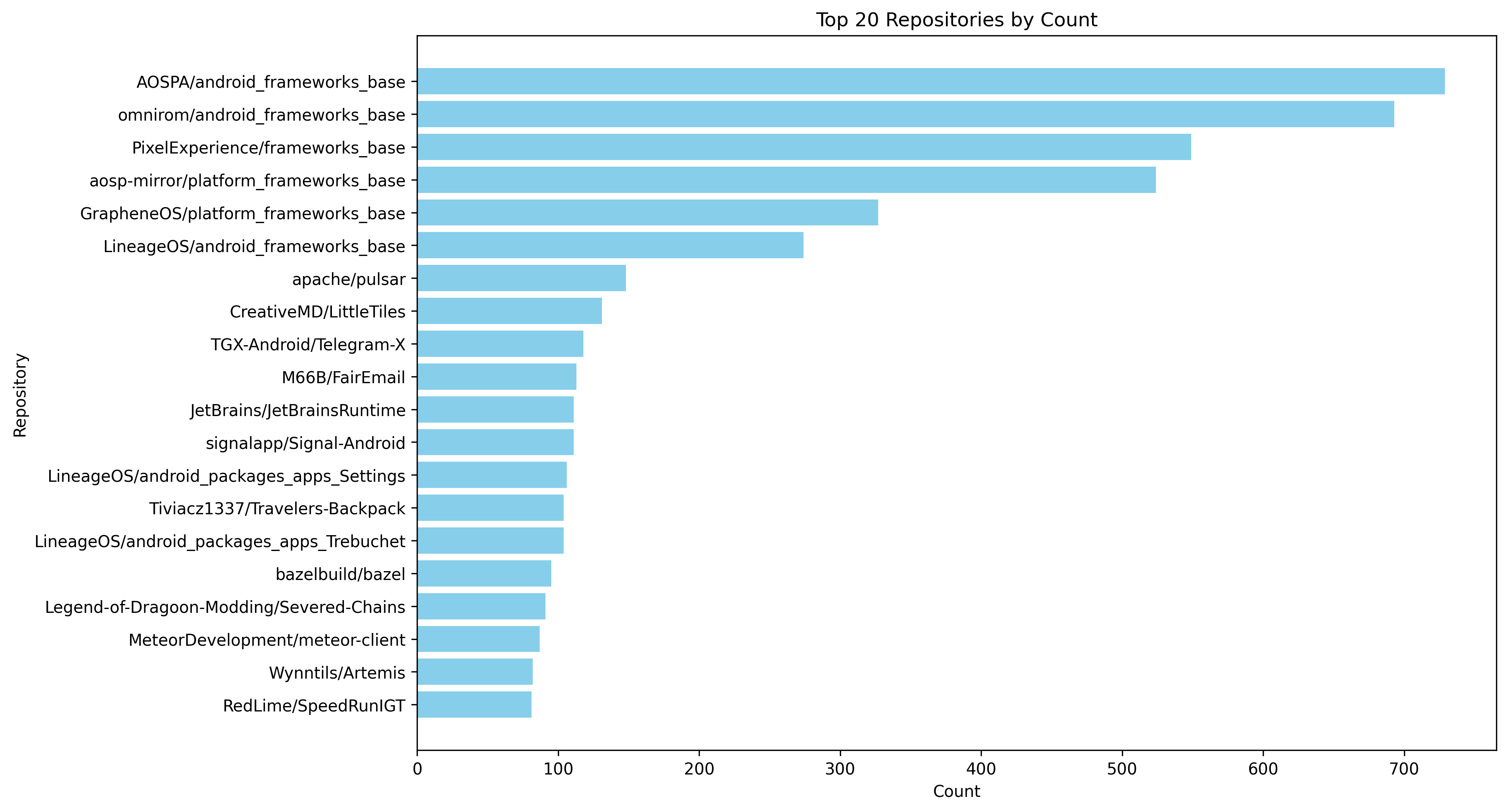}
    \caption{Top 20 GitHub repositories with the highest contributions to \manualdataset and \autodataset}
    \label{fig:top20}
\end{figure*}

The two variants, \manualdataset and \autodataset, differ in their scale and method of collection. The \manualdataset dataset, which was manually curated, includes data from 263 projects with a total of 784 unique code commits. This dataset represents a smaller but highly accurate collection, offering a focused selection of VFCs that were carefully verified and annotated by experts. In contrast, the \autodataset dataset was generated using an automated approach, resulting in a much larger collection of 16,837 code commits across 2,532 projects. While broader in scope, this dataset benefits from automated processes that ensure a comprehensive capture of VFCs from a wide range of Java OSS projects.

To further illustrate the distribution of VFCs across repositories, Figure \ref{fig:top20} shows the number of commits containing fixes in the top 20 GitHub repositories within the \manualdataset and \autodataset datasets. The repository with the highest number of commits is AOSPA/android\_frameworks\_base, with 729 commits, followed closely by omnirom/android\_frameworks\_base and PixelExperience/frameworks\_base, with 693 and 621 commits, respectively. These repositories, along with three others focused on frameworks base, collectively account for 18.2\% of the total commits in the datasets. The top 20 repositories contribute 4,658 commits, representing 26.4\% of the total commits, highlighting the concentration of VFCs in certain high-profile projects.

% This dataset is a valuable resource for researchers interested in security and vulnerability prediction, providing a robust foundation for analyzing the patterns and characteristics of VFCs in Java OSS projects on GitHub.

\section{Dataset Application}
% Security vulnerabilities in software are a critical concern for developers and organizations alike. Addressing these challenges requires advancements in areas such as vulnerability detection, vulnerability fixing commits detection, and code quality. 
Our dataset allows future research in several areas, including but not limited to:

\textbf{VFC Detection:} This dataset enables research focused on identifying commits that fix vulnerabilities. By analyzing historical commits and their metadata, researchers can develop methods to recognize potential VFCs based on code changes, commit messages, and project characteristics. 
These models could be integrated into continuous integration pipelines.
This would aid developers in identifying vulnerable code and improving the overall security and stability of software projects.

\textbf{Vulnerability Detection:} While our dataset does not directly contain vulnerability-inducing information, it serves as a valuable starting point for software vulnerability detection tasks, 
including the investigation into future research for the use of SZZ~\cite{sliwerski2005changes}. %\textcolor{red}{add some references here?}
Future research can also use this dataset to extract vulnerability information, including code at different granularities, such as file level or function level.

\textbf{Vulnerability Repair:} 
% Each VFC addresses a vulnerability. 
From the version of the code fixed by the VFC, we can obtain the vulnerable code, while the version of the code after the VFC gives us the fixed version of the code. 
Our dataset is also suitable for research on vulnerability repair as it demonstrates how developers fix vulnerable code.

\textbf{Empirical Studies:} The dataset provides a rich resource for conducting empirical studies. Researchers analyze the Common Weakness Enumeration (CWE) categories into which the vulnerabilities fall. 
This would offer insights into prevalent security issues. 
Such studies could inform best practices in secure coding and guide the development of more targeted security tools and techniques.

% The dataset facilitates empirical studies on the process of fixing vulnerabilities in software. Researchers can analyze the methods and approaches used by developers to fix vulnerabilities, identifying common patterns and best practices. These studies could also investigate the impact of developer expertise and project characteristics on the effectiveness of vulnerability fixes, providing insights that could improve the training and support provided to developers.

\section{Threats to Validity}

\textbf{Threats to Internal Validity.} The process of identifying VFCs relies heavily on keyword-based searches within commit messages. 
This approach is effective for processing our large dataset of 34,321 Java projects, 
but may miss commits that address vulnerabilities but do not explicitly use the selected keywords. Conversely, it may also include commits that are not actually related to vulnerability fixes even if they match the keywords.
While this is a threat, we provide a variant of the dataset where each instance is manually validated by human annotators when a scenario requires a high precision.
% we achieved a precision of 0.7.
% We acknowledge these limitations and plan to explore alternative VFC classification methods in future work to enhance the accuracy of our identification process.

\textbf{Threats to Construct Validity.} To establish the final set of keywords used for filtering, we conducted several rounds of annotation. Although this process involved trained annotators and considerable efforts were made to achieve a high level of agreement, the potential for human error and subjectivity remains. Differences in interpretation among annotators could lead to inconsistencies in the dataset. 
However, we believe this threat is minimal, given that we achieved a precision of 0.7 and a Fleiss' Kappa score of 0.7, which indicates substantial agreement among the annotators.

\textbf{Threats to External Validity.} Our dataset focuses exclusively on Java projects, which may limit the generalizability of our findings to other programming languages. Certain types of vulnerabilities might be underrepresented, particularly those that are less commonly addressed or reported in Java projects. This language-specific focus could potentially skew our understanding of vulnerability patterns across different programming paradigms. To address this limitation and enhance the external validity of our findings, we plan to extend this dataset in future work to include vulnerabilities from other programming languages.

% \textbf{Scalability and Generalization.} The methodologies and tools used in curating the dataset were used on Java projects. As a result, the findings and insights drawn from this dataset may not directly generalize to other programming languages or platforms without further adaptation and validation.

\section{Conclusion}
% VFCs are valuable but often limited resources. 
% Existing Java VFC datasets heavily rely on NVD and project-specific web resources.
VFCs are invaluable resources for software security research, yet they are often scarce. To address this limitation and support Java vulnerability analysis, we present a comprehensive VFC dataset with two distinct variants: \manualdataset and \autodataset. The \manualdataset comprises 784 VFCs meticulously verified by expert annotators, ensuring high reliability. In contrast, the \autodataset offers a broader scope with 16,837 VFCs, identified through keyword-based filtering of commit messages. This dual-variant approach balances precision and scale, catering to diverse research needs. Our dataset facilitates various research avenues, including VFC detection, vulnerability identification, automated vulnerability repair, and beyond. To ensure accessibility and ease of use, we have made the dataset available in JSONL format, enabling seamless integration into research workflows.

% \newpage

\bibliographystyle{IEEEtran}
\bibliography{main}

\end{document}